# Beam Blockage in Optical Wireless Systems

Sarah O. M. Saeed[1], Sanaa Hamid Mohamed[1], Osama Zwaid Alsulami[1], Mohammed T. Alresheedi[2], Taisir E. H. Elgorashi[1] and Jaafar M. H. Elmirghani[1]

[1]*School of Electronic and Electrical Engineering, University of Leeds, LS2 9JT, U.K.*
[2]*Department of Electrical Engineering, King Saud University, Riyadh, Kingdom of Saudi Arabia*
elsoms@leeds.ac.uk, elshm@leeds.ac.uk, ml15ozma@leeds.ac.uk, malresheedi@ksu.edu.sa,
t.e.h.elgorashi@leeds.ac.uk, j.m.h.elmirghani@leeds.ac.uk

**ABSTRACT**

In this paper, we use the percentage blockage as a metric when an opaque disc obstructs the Line-of-Sight link from the access point to the receiver in an optical wireless indoor communication system. The effect of the different parameters of the obstructing object are studied, these are the radius, the height, and the horizontal distance from the receiver in the positive y direction. The percentage of blocked room locations to the total number of room locations when varying the disc parameters is studied assuming a single serving link. It was found that depending on the dimensions of the obstructing object and the distance from the receiver in addition to which access point is serving the user, that blockage can vary between 0% up to 100%. Furthermore, the service received by a user, in terms of beam blockage depends on the access point they are connected to. The resulting fairness challenges will be addressed in resource allocation optimization in future work.

## 1. INTRODUCTION

Optical Wireless (OW) communication is a promising technology that can complement the radio frequency (RF) based communication systems and can help meet the exponentially increasing demand for data. Both OW and RF based systems have their advantages and disadvantages in relation to spectrum and channel characteristics. While the inability of light to penetrate opaque objects is considered a merit exploitable to provide physical layer security and increased capacity by the reuse of optical frequencies in other rooms in OW systems (as opposed to RF based systems), it is considered as a disadvantage if an object blocks the direct line-of-sight (LOS) beam within a room. Higher order reflections from different room surfaces can be utilized, however, the LOS beam has the highest contribution to the received optical power compared to n-order reflections that undergo diffusion and losses due to absorption by reflecting surface materials [1]. In addition, the multipath propagation due to optical rays arriving from reflecting surfaces results in pulse dispersion, creating inter-symbol-interference (ISI) and hence limiting the achievable data rate. Multi-gigabit per second OW systems were reported in literature [2]–[6]. These systems enhance the achievable data rate by improving the received power, reducing the effects of background noise, and the pulse delay spread employing different techniques. These techniques; such as multi-spot diffusing transmitters, diversity transmitters and receivers [7], and beam delay and power adaptation techniques [8], [9] with beam steering and Computer Generated Holograms (CGH) [10], [11]; aim at enhancing the received optical power by enhancing the LOS component of the received optical power and reducing the noise, which result in higher signal-to-noise ratio (SNR) and hence higher data rates.
Ensuring the availability of a direct LOS link and the resilience against blockage is important for a reliable high data rate system that meets the service level agreement (SLA) requirements.
In [12], shadowing and blockage were studied for the cellular and multispot diffusing MIMO systems. In [13], WiFi was proposed to provide an alternative service in case of optical link blockage. In this work a preliminary study of blockage is performed for a simplified object that can be later combined with other complex objects to represent real objects in the environment. Here the effects of the blockage of the LOS links by disc objects within an indoor environment is studied. The percentage of room locations at which the beam is blocked is calculated assuming a single serving access point (AP). The rest of this paper is organized as follows: Section 2 describes the procedure used to evaluate the percentage blockage under different obstructing object parameters. Section 3 shows the results obtained. Conclusions and future work are provided in Section 4.

## 2. LINE-OF-SIGHT BEAM BLOCKAGE BY A DISC

In this work, the effect of blockage of LOS links by objects within indoor room environment is studied. The room considered has dimensions of (4 m x 8 m x 3 m) for (width x length x height) respectively. Receivers are assumed to be distributed over the communication floor (CF) at 1 m above ground which is the height of a typical office desk. Eight sources of illumination (or access points) to provide sufficient illumination [4] are assumed at the (x,y,z) coordinates of (1,1,3), (1,3,3), (1,5,3), (1,7,3), (3,1,3), (3,3,3), (3,5,3) and (3,7,3) as shown in Fig. 1. The APs serve the receivers based on fixed allocation by the controller [14]. A horizontal disc with radius $R$, height $h$ effectively calculated above the CF, and horizontal distance $d$ between the receiver and the disc centre in the positive $y$ direction, is assumed. The disc can be generalized and more complexity can be added so that it represents other realistic objects in the room (whether mobile or fixed), such as a human, a paper held by a human, or a piece of furniture.

Assuming a single receiver in the room assigned a single AP, two parameters of the disc are fixed, and the third parameter is varied as shown in Fig. 2. For specific $R$, $h$, and $d$, the receiver location over the room is varied every 25 cm all over the CF. The percentage blockage is then calculated as the percentage of the number of blocked locations to the total number of locations.

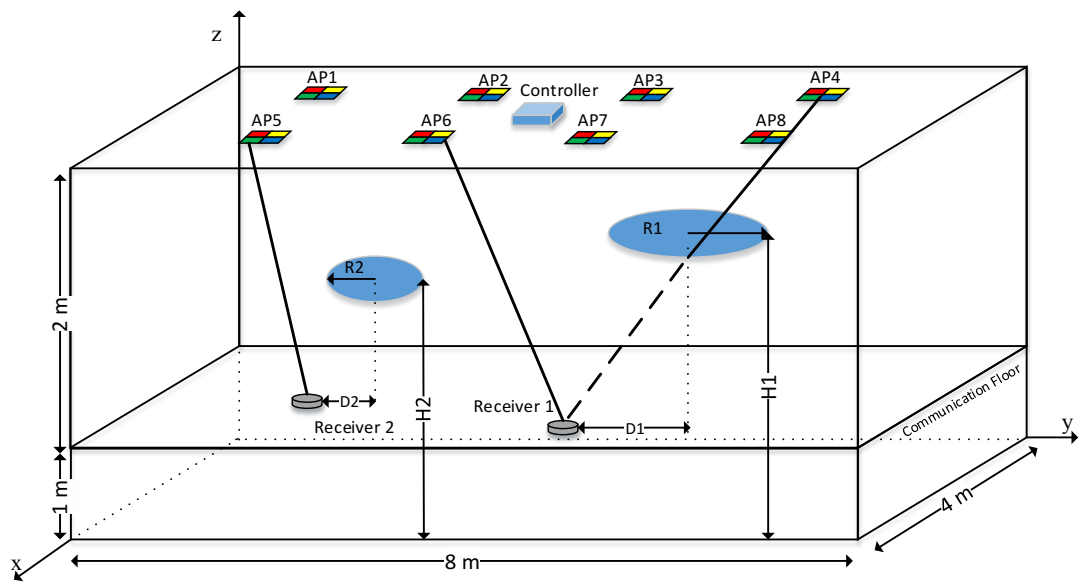

*Figure 1. Room model with two receivers. Receiver 1, which is served by AP 4, is blocked by Object 1 while Receiver 2 that is served by AP 5 is not blocked for the scenario illustrated. AP 6 provides a protection link to Receiver 1. Note that if both Receiver 1 and 2 move to the right, their blockage status can change.*

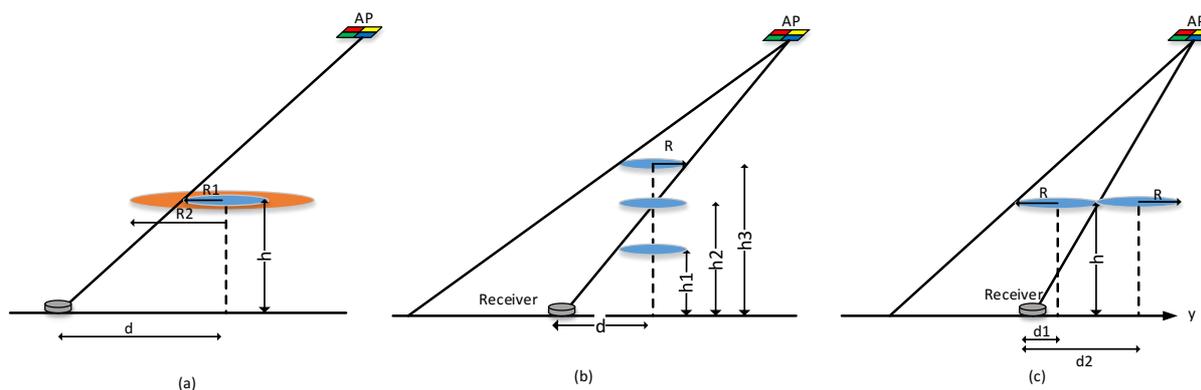

*Figure 2. Illustration of the effect of varying the disc parameters. (a) the beam passes over the disc when the radius is R1 or less than R1. Increasing the disc radius more than that leads to LOS beam blockage. (b) the LOS beam passes over the lowest disc with height h1, but the beam is blocked if the height is further increased until the beams passes below the disc if it is h3. (c) the beam passes below the disc at distance d1 and when the disc moves in the positive y direction, it is blocked until it is at distance d2 where it passes above the disc.*

## 3. RESULTS AND DISCUSSION

It can be seen that in Fig. 3 that one disc parameter is varied while the other two parameters are fixed at a minimum value, a maximum value, and a value in between.

The results obtained are shown in Fig. 3 when varying the values of $R$, $h$, and $d$. In Fig. 3 (a-c) where $R$ is varied, it can be seen that increasing the radius of the disc leads to increased percentage of blockage. For a given value of $h$ and $d$, if $R$ is increased, blockage increases. This can be expected as shown in Fig. 2 (a) where the beam is blocked after a specific value of R for a given AP and receiver location and given $h$ and $d$ values.

In Fig. 3 (d-f) where $h$ is varied, it can be seen that if the height is zero above the communication floor and the radius of the disc is greater than or equal to the horizontal distance between the disc centre and the receiver, the receiver is always blocked as it is covered by the disc, otherwise there is no blockage at $h$ equal to zero. In addition to the relation between $R$ and $d$ and since the disc is assumed at a horizontal distance $d$ in the positive $y$-axis direction and depending on whether the location of the assigned AP is to the left or right of the receiver, the probability of blocking varies over the whole CF. Given $R$ is less than $d$ and assuming the disc lies in between the receiver and AP (across a horizontal separation), and as illustrated by Fig. 2 (b) the percentage beam blockage starts from 0 when the LOS beam passes over the disk, and increases to some value and then decreases, when the beam passes under the disc. It can be seen that if $R$ is greater than $d$, the beam blockage percentage starts from 1 and then decreases.

The results when varying $d$ are shown in Fig. 3 (g-i). These results experience a similar pattern to the previous case when $h$ was varied. This is also interpreted using Fig. 2(c); when $d$ increases, the beam passes under the disc, gets blocked, and then passes over the disc. After the distance between the receiver and the obstructing object exceeds some value depending on the separation, the object no longer blocks the receiver and the percentage beam blockage goes to zero.

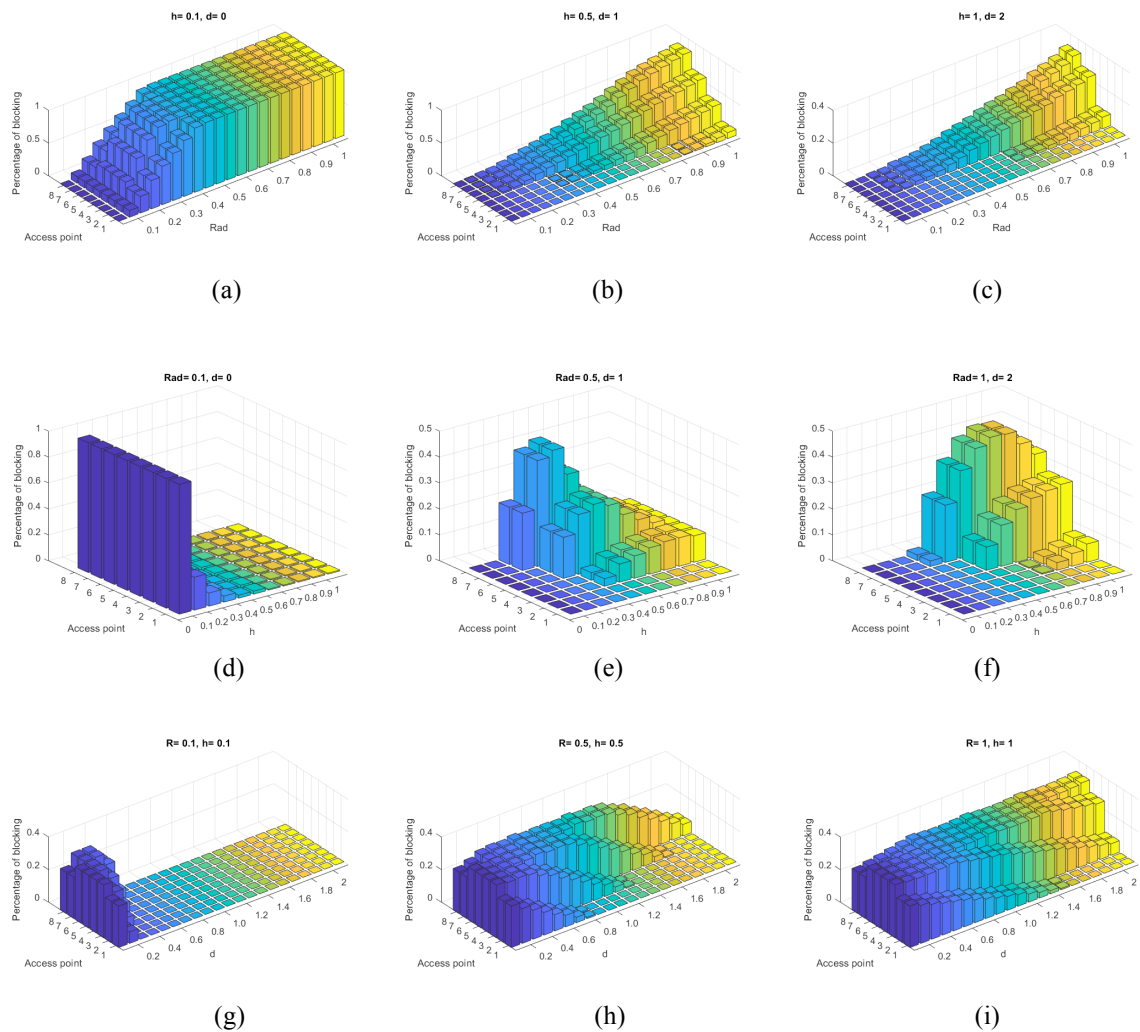

*Figure 3. Blocking probability varying one disc parameter and fixing the other two. The different APs in the room are considered.*

## 4. CONCLUSIONS AND FUTURE WORK

In this paper, the effect of direct line-of-sight beam blockage by an opaque object was studied. The blockage is studied for different values of the parameters of the obstructing object including the radius, height, and horizontal distance from the receiver; where blockage is measured using the percentage of beams blocked. In our planned future work, implementing resilience by assigning two or more links is expected to reduce blockage on average. The effect of the added traffic as a result of using redundant links, and its impact on the network can also be studied in the future.

In this paper only blockage of LOS links was considered, however, shadowing of higher order reflection components needs to be studies to evaluate the ability to work under a shadowing object for example. Also the problem of blockage can be examined to obtain the statistical distribution of the link blockage probability and this can be obtained both through simulations and measurements and can be used so that allocation of access points is blockage-aware in static environments while in dynamic environments, allocation can be handled using intelligent MAC protocols.

**ACKNOWLEDGEMENTS**

The authors would like to acknowledge funding from the Engineering and Physical Sciences Research Council (EPSRC) for the TOWS project (EP/S016570/1). All data are provided in full in the results section of this paper.